\documentclass{elsart}
\usepackage{amsmath}
\usepackage{amssymb}

\usepackage{bm}
\usepackage{graphicx}
\usepackage{epsfig}
\usepackage{hyperref}
\usepackage[english]{babel}

\begin{document}

\begin{frontmatter}

%%%%%%%%%%%%%%%%%%%%%%%%%%%%
\title{Competing effective interactions of Dirac electrons in the AF Heisenberg-Kondo lattice }
%%%%%%%%%%%%%%%%%%%%%%%%%%%%

%%%%%%%%%%%%%%%%%%%%%%%%%%%%
%
\author[UFRJ]{E. C. Marino}
\ead{marino@if.ufrj.br}
and
\author[UFSJ]{Lizardo H. C. M. Nunes}
\ead{lizardonunes@ufsj.edu.br}

\address[UFRJ]
{Instituto de F\'{\i}sica, Universidade Federal do Rio de Janeiro, Caixa Postal 68528, Rio de Janeiro, RJ, 21941-972, Brazil}

\address[UFSJ]
{Departamento de Ci\^encias Naturais, Universidade Federal de S\~ao Jo\~ao del Rei, 36301-000 S\~ao Jo\~ao del Rei, MG, Brazil}

%%%%%%%%%%%%%%%%%%%%%%%%%%%%

%%%%%%%%%%%%%%%%%%%%%%%%%%%%
\begin{abstract}
Recently discovered advanced materials frequently exhibit a rich phase diagram suggesting the presence of different competing
interactions. A unified description of the origin of these multiple interactions, albeit very important for the comprehension of such materials is, in general not available. It would be therefore very useful to have a simple model where the common source of different interactions could be possibly traced back.
 In this work we consider the (AF) Heisenberg-Kondo lattice model, describing a set of localized spins on a square lattice with anti-ferromagnetic nearest neighbors interactions and itinerant electrons, which are assumed to be Dirac-like. These interact with the localized spins through a Kondo magnetic interaction. By integrating out the localized degrees of freedom we obtain a set of different effective interactions among the itinerant electrons. This includes a BCS-like superconducting term, a Nambu-Jona-Lasinio-like, excitonic term and a spin-spin magnetic term. The resulting phase diagram is investigated by evaluation of the mean-field free-energy as a function of the relevant order parameters. This shows the competition of the above interactions, depending on the temperature, chemical potential and coupling constants.
\end{abstract}
%%%%%%%%%%%%%%%%%%%%%%%%%%%%

%%%%%%%%%%%%%%%%%%%%%%%%

%%%%%%%%%%%%%%%%%%%%%%%

\end{frontmatter}

%%%%%%%%%%%%%%%%%%%%%%%%%%%%
%Beginning of the Text
%%%%%%%%%%%%%%%%%%%%%%%%%%%%
\section{Introduction}\label{Introduction}

Many advanced materials, which were discovered recently, exhibit a phase diagram so rich that suggests the presence of different competing interactions as the reason behind the various types of ordering. Abundant experimental data exist for several materials. Nevertheless, a clear unifying picture that would allow the understanding of the detailed mechanisms that generate such competing interactions is not yet available. Such a picture, however, could be quite well one of the key requirements for understanding the essential physics of these materials. It could shed light, for instance, on the nature of the
mechanisms that produce the onset of high-Tc superconductivity in cuprates and iron pnictides.

In many new materials we find a phase diagram where we can observe, for instance,
a deep interplay among
 superconducting, magnetic and charge orderings.
 Some of these materials have attracted a lot of attention recently, among them high-Tc superconductors such as the
 cuprates and  iron pnictides and low-Tc superconductors such as the transition metal dichalcogenides (TMD). Graphene
can also display a rich phase diagram, provided
 it is doped, strained or has adsorbed atoms on its surface~\cite{graphene}.

Interestingly, for all of the systems mentioned above it happens that, under different conditions and for different reasons,
the electronic excitations behave as Dirac fermions.
%These include iron pnictides~\cite{pnictides},
Iron based pnictide materials, for instance,
undergo a transition from a magnetically ordered state to a superconducting one
upon doping~\cite{Cruz2008,Kamihara2008,Rotter2008}.
For the particular case of the 122 materials,
magnetic order and superconductivity coexist in a small region of the phase diagram~\cite{Goko2009}
and the new quasiparticles in that region
exhibit a Dirac-like linear energy dispersion relation~\cite{Richard2010,Khuong2011,PRL2011}

%superconducting cuprates~\cite{cuprates},
In the case of cuprate superconductors the parent compounds  are insulators presenting  AF order.
As charge carriers are added to the CuO$_{ 2 } $ planes, there is the onset of superconductivity,
with the characteristic dome-shaped superconducting phase diagram~\cite{cuprates,Nagaosa2006}.
Dirac points appear in the intersection of the nodes of the $ d $-wave superconducting gap and the two-dimensional (2D) Fermi surface~\cite{Affleck1988,Affleck1989,Wen1996}.
Strongly interacting 2D Dirac fermion systems also exhibit a dome structure in their superconducting phase diagram~\cite{Smith2009,Nunes2010}, so
we might inquire about the possibility of Dirac fermions playing some role in the description of the cuprate superconductors.

%transition metal dichalcogenides
Moroever, the quasi 2D TMD are layered compounds
where $ s $-wave superconductivity coexists with a charge density wave (CDW) at low temperatures
and applied pressure~\cite{Withers1986,Wilson1969,Wilson1975}.
A theory has been proposed in which Dirac electrons appear close to the nodes of the CDW gap
and form Cooper pairs~\cite{CastroNeto2001,Uchoa2005}.
The theory is consistent with the linear decay of the temperature dependent critical field~\cite{Marino2007},
which is observed experimentally in the copper doped dichalcogenide Cu$_x$TiSe$_2$~\cite{Morosan2006}.

%graphene
Graphene, on the other hand, is a semi-metal with gapless electronic quasiparticles, which
due to the peculiar lattice structure behave as Dirac fermions. Pure graphene does not exhibit
superconductivity or magnetism,
however, upon doping and/or straining may display competing orders, such as local magnetic moments, superconductivity or an excitonic gap~\cite{Kotov2011}.

Heavy fermions are another vast class of materials presenting a rich competition of different types of order in their phase diagrams~\cite{heavyfermions,Stewart2001,Stewart2006}

The above mentioned systems have been
extensively studied both theoretically and experimentally but, nonetheless, we still do not have a clear unified picture affording
a detailed understanding of the microscopic mechanisms that lead to each kind of ordering in the phase diagram.
It would be very instructive and inspiring,
therefore, to find fully controllable models for systems displaying phases with the
aforementioned different types of ordering, where one could precisely trace back the original interaction and the mechanisms leading to such phases.

 In this work, we investigate the AF Heisenberg-Kondo lattice model~\cite{spin-fermion},
 a system containing both localized spins and itinerant electrons with a mutual Kondo-like magnetic interaction.
Our main aim is to determine what is the effective net interaction among the conduction electrons,
which results from their primary magnetic interaction with the AF substrate of localized spins. For this purpose, we use the well-known
 Nonlinear Sigma Model (NLSM) description of the latter~\cite{Chakravarty89}.
The itinerant electrons, conversely, are supposed to have a tight-binding band structure
showing the presence of Dirac cones whose vertices touch at the interface between the valence and conduction bands.
We assume the system to be close to half-filling and therefore describe the kinematics of the itinerant electrons by the Dirac hamiltonian.
In order to describe the the magnetic Kondo interaction it is convenient to re-phrase the NLSM in the CP$^1$ language~\cite{Auerbach94},
whereby that interaction becomes a quartic term involving two Dirac fermion fields and two bosonic (Schwinger Boson) fields.

Our strategy will be to functionally integrate over the bosonic fields in order to derive the resulting effective interaction among the fermion fields. In this process, we show that the Kondo interaction among itinerant and localized electrons can be completely expressed
as a gauge coupling between the Dirac fermions and the Schwinger bosons, mediated by the CP$^1$ vector gauge field.
Our final result for the effective electron interaction is then obtained upon integration over this gauge field.
The resulting interaction possesses three pieces: a superconducting BCS-type term,
an Ising-like magnetic interaction
and a Nambu-Jona-Lasinio-type interaction.
These terms will favor, respectively, superconducting, magnetic and insulating charge-gapped orderings.
Our main result is the demonstration that
the original system of localized spins and itinerant electrons with mutual magnetic interaction produces, ultimately, an interacting electronic system, which presents a variety of competing orders similar to the ones
observed in the materials mentioned above.

We do not pretend here to be modeling those complex  materials with the  AF Heisenberg-Kondo lattice model. Nevertheless, we provide an
example of a simple system presenting multiple possible orderings, which can be ascribed to different effective interactions, all derived
from the same original magnetic interaction among localized spins and itinerant electrons.

We complete our study by investigating the phase diagram produced by the effective interactions we found in this model.
We do this by performing a mean-field calculation of  the finite-temperature free-energy (effective potential) as a function of the different
order parameters. Subsequently  we analyze the free-energy minima, and out of these, determine the value of the
different order parameters as a function of temperature and coupling constants.

%%%%%%%%%%%%%%%%%%%%%%%%%%%%
\section{The Model and Its Continuum Limit}\label{TheModel}

We consider a single layered system containing both localized and itinerant spins in which the former are located at
the sites of a square lattice and have an antiferromagnetic exchange integral while the latter are conduction electrons
with a tight-binding dispersion relation, which is assumed to be Dirac-like. The localized spins mutual interaction will be described
by an AF Heisenberg hamiltonian on a 2D square lattice, whereas their interaction with the itinerant ones, by a Kondo-like term. The complete
hamiltonian, therefore, contains three parts, namely
\begin{eqnarray}
 H
%& = &
=
J \sum_{ \langle ij \rangle } {\bf S }_i \cdot {\bf S }_j
- t \sum_{ \langle ij \rangle } \left ( c_{i\alpha}^\dagger c_{j\alpha} + hc  \right)
%\nonumber \\
%& &
+ J_K \sum_{i} {\bf S }_i \cdot \left ( c_{i\alpha}^\dagger {\vec \sigma}_{\alpha\beta} c_{i\beta} \right )
\, ,
\label{EqH}
\end{eqnarray}
where ${\bf S }_i$ is the localized spin operator and $c_{i\alpha}^\dagger$ is the creation operator of an itinerant electron of spin $\alpha=\uparrow,\downarrow$, both at site $ i $. Frequently we have materials for which there are electrons coming from different bands
or even from inequivalent regions of the Brillouin zone. In these cases we would add an extra label $a=1,...,N$ to the electron operators.
Here, for the sake of simplicity, we shall omit such a label, this fact having no effect in our conclusions.

In order to obtain the partition function,
we employ the continuum path integral approach. By using a basis of spin coherent states we have the localized spin operators $ {\bf S }_i$ replaced by
 $S {\bf N }( {\bf x } ) $, where $S$ is the spin quantum number and  $  {\bf N }( {\bf x } )  $ is a classical vector such that
 $ | {\bf N }( {\bf x } ) |^{ 2 } = 1 $.  $ {\bf N }$ is then decomposed into two perpendicular components, ${ \bf L}$ and ${ \bf n}$,
 ($ {\bf L } \cdot { \bf n} = 0 $) associated respectively
with ferromagnetic and antiferromagnetic fluctuations\cite{sachdev,Tsvelik95},
\begin{equation}
{\bf N }( {\bf x } ) = a^2 { \bf L}( {\bf x } ) + ( - 1 )^{ | {\bf x} | } \sqrt{ 1- a^{4}| {\bf L } |^{2} } \, { \bf n}( {\bf x } )
\, ,
\label{EqCoherentDecomposition}
\end{equation}
where  $ a $ is the lattice parameter. In the continuum limit ($a\rightarrow 0$), this becomes
\begin{equation}
{\bf N }( {\bf x } ) = a^2 { \bf L}( {\bf x } ) + ( - 1 )^{ | {\bf x} | }  \, { \bf n}( {\bf x } ) + O(a^4)
\, .
\label{EqCoherentDecomposition1}
\end{equation}
Notice that we always have $ | {\bf n }( {\bf x } ) |^{ 2 } = 1 $.

In terms of these and of the continuum fermion field  $ \psi_{ \alpha }( {\bf x } ) $ corresponding to $c_{i\alpha}$
we can express the partition function as the functional integral
\begin{eqnarray}
\mathcal{Z}
& = &
\int
\mathcal{D}\psi
\,
\mathcal{D}\psi^{ \dagger}
\,
\mathcal{D}{\bf L}
\,
\mathcal{D}{\bf n} \ \delta\left[ | {\bf n } |^{ 2 } - 1 \right]
 \nonumber \\
& &
\times
\exp
\left[ - \int_0^\beta d\tau \int d^2 x \,
\left( \,
\mathcal{ H }
-
\psi^{ \dagger } i\partial_{ \tau } \psi
\, \right)
\right]
\, ,
\label{EqZ}
\end{eqnarray}
where the continuum hamiltonian density reads
\begin{eqnarray}
{\mathcal H }
& =  &
\psi^{ \dagger } \left( i \vec \sigma \cdot \vec \nabla - \mu \right)   \psi
+ \rho_{ s } | {\bf \nabla } {\bf n } |^{ 2 } + \chi_{ \perp } S^{2 } | {\bf L } |^{ 2 }
\nonumber \\
& &
+
S a^2 {\bf L } \cdot \left[ J_{ \mbox{\scriptsize{K}} }  \; {\bf s }
+
i a^{-2} \, \left(  {\bf n } \times \partial_{ \tau } { \bf n}  \right)\right]
%\nonumber \\
%& &
+ ( - 1 )^{ | {\bf x} | } S  J_{ \mbox{\scriptsize{K}}}{\bf n } \cdot {\bf s }
\, ,
\label{EqHNLSM}
\end{eqnarray}
with $ \rho_{ s } = J S^{ 2 } $ being the spin stiffness, $ \chi_{ \perp } = 4 J a^2$ the transverse susceptibility
and $ {\bf s } $  the itinerant electron spin operator, given by
\begin{equation}
{\bf s }
=
 \psi^{ \dagger }_{ \delta } \left(  \vec{ \sigma } \right)_{ \delta \gamma } \psi_{ \gamma }
\, .
\label{Eqs}
\end{equation}

In  expression (\ref{EqHNLSM}), the first term is the continuum electron kinetic hamiltonian density, derived from the tight-binding
energy assuming the system has a Dirac-like dispersion relation near the Fermi points.
As the system is doped, charge carriers are added or removed from the conduction band. Their total number
is controlled by a chemical potential $\mu$, which has, therefore, been included in the previous equation.
The fourth and sixth terms correspond to the original Kondo interaction term and the remaining terms in (\ref{EqHNLSM})
are the usual ones associated to the AF Heisenberg term \cite{sachdev}.

Integrating over $ \bf L $ in (\ref{EqZ})
we obtain the resulting effective lagrangian density
\begin{eqnarray}
\mathcal{ L }_{  \mbox{\scriptsize{eff}}}
& = &
 \psi^{ \dagger } \left[ i\gamma^0\gamma^\mu\partial_\mu -\mu\right] \psi
+
\frac{ \rho_{ s }  }{ 2 }
\left(
| {\bf \nabla } {\bf n } |^{ 2 } - \frac{ 1 }{ c^{2 } } | \partial_{ t} {\bf n } |^{ 2 }
\right)
\nonumber \\
& &
\hspace{-0.7cm}
+
a^2 J_{ \mbox{\scriptsize{K}} }
\left\{( - 1 )^{ | {\bf x} | } S a^{-2} {\bf n } +
\frac{ i }{ \chi_{ \perp} } \left(  {\bf n } \times \partial_{ \tau } {\bf n } \right)
+  \frac{a^2 J_{ \mbox{\scriptsize{K}} } }{ 2 \chi_{ \perp} } {\bf  s }
\right\}
\cdot {\bf  s }
\, ,
\label{EqH2}
\end{eqnarray}
 where $(\gamma^0)^2=1$, $\gamma^0\gamma^i=\sigma^i$ and
 $ c = \sqrt{ \rho_{ s } \chi_{ \perp } } $ is the spin-wave velocity.

It will be convenient to use the CP$^1$ (Schwinger Boson) formulation of the
 O(3) NLSM, in which the AF fluctuation field is written as
\begin{equation}
n_{ i } =  z^{ * }_{ \alpha } \left( \sigma_{ i } \right)_{ \alpha \beta } z_{ \beta } \, , \ \ \ \  i = x, y, z,
\,
\label{Eqz1z2}
\end{equation}
in terms of the two complex fields $ z_{ \alpha }$, $\alpha = 1, 2 $,
satisfying the constraint
 $ | z_{ 1} |^{ 2 } + | z_{ 2 } |^{ 2 } = 1 $. In the CP$^1$ language the effective lagrangian density
 (\ref{EqH2}) is rewritten as
\begin{eqnarray}
\mathcal{ L }_{  \mbox{\scriptsize{eff}}}
& = &
 \psi^{ \dagger }  \left[ i\gamma^0\gamma^\mu\partial_\mu -\mu\right] \psi
+
2 \rho_{ s } |  D_\mu z_i  |^2
\nonumber \\
& &
+
a^2 J_{ \mbox{\scriptsize{K}} }  \  \psi^{ \dagger } \left [ ( - 1 )^{ | {\bf x} | }S   a^{-2} \  \vec{ \sigma }\cdot {\bf n }
+ \frac{ i }{ \chi_{ \perp} } \vec{ \sigma }\cdot \left( {\bf n } \times \partial_{ \tau } {\bf n } \right)\right ]
\psi
%\nonumber \\
%& &
+  \frac{a^4  J_{ \mbox{\scriptsize{K}} }^2 }{ 2 \chi_{ \perp} } {\bf  s}
\cdot {\bf  s }
\, ,
\label{EqH21}
\end{eqnarray}
where the components of ${\bf n}$ are given by (\ref{Eqz1z2}) and
$D_\mu = \partial_\mu-i A_\mu$.

%%%%%%%%%%%%%%%%%%%%%%%%%%%%
\section{ Replacing the Magnetic Interaction by a Gauge Coupling}

We now perform a canonical transformation \cite{Marino2002,Kubert1993} on the electron field, namely
\begin{equation}
\psi_{ \alpha } \rightarrow  U_{ \alpha \beta } \, \psi_{ \beta }
\, ,
\label{EqCanonicalTransformation}
\end{equation}
where the unitary matrix $ U $ is written in terms of the $ z_{ \alpha } $-fields as
\begin{equation}
U
=
\begin{pmatrix}
  z_{ 1 }   &  - z^{ * }_{ 2 }  \\
  z_{ 2 }   &   \ \  z^{ * }_{ 1}
\end{pmatrix}
\, .
\label{EqU}
\end{equation}

This matrix has the following property
\begin{equation}
U^\dagger \vec{ \sigma }\cdot {\bf n } U = \sigma^z
\label{EqH04}
\end{equation}
and therefore the first term in the second line in (\ref{EqH21}) can be expressed, up to a sign, as the density difference of electrons
with opposite spins:
\begin{equation}
( - 1 )^{ | {\bf x} | } J_{ \mbox{\scriptsize{K}} }  S  \left( \psi^\dagger_{\uparrow } \psi_{\uparrow }  -\psi^\dagger_{\downarrow } \psi_{\downarrow }
\right).
\label{1}
\end{equation}

Assuming a uniform density of electrons, we conclude that this term will vanish upon integration in  ${\bf x}$ because of the
rapidly oscillating pre-factor.

Now, since $U$ represents a local operation,
it follows that, under the transformation (\ref{EqCanonicalTransformation}), the electron kinetic term generates
the additional interaction
\begin{equation}
i \  \psi^{ \dagger }  \gamma^0\gamma^\mu \left (U^\dagger \partial_\mu U \right) \psi
\, .
\label{2}
\end{equation}

From (\ref{EqU}),
we obtain
\begin{eqnarray}
U^\dagger \partial_\mu U
=
i \sigma^z \ A_\mu
%\nonumber \\
%& &
%\hspace{-2.7cm}
+
\begin{pmatrix}
0 &  z^*_{ 2 } \, \partial_{ \mu } z^{ * }_{ 1 } -  z^*_{ 1 } \, \partial_{ \mu } z^{ * }_{ 2 } \\
- z_{ 2 } \, \partial_{ \mu } z_{ 1 } +  z_{ 1 } \, \partial_{ \mu } z_{ 2 }   &   0
\end{pmatrix}
\, ,
\label{EqA}
\end{eqnarray}
where we used the fact that $A_\mu = -i z^{ * }_{ i } \, \partial_{ \mu } z_{ i }$ , which follows from (\ref{EqH21}).

Now, consider the polar representations of the fields $ z_{ \alpha } $,
\begin{equation}
z_{ \alpha } = \frac{ \rho_{ \alpha } }{ \sqrt{ 2 } } \, e^{ i \, \theta_{ \alpha }  }
\, , \ \ \ \  \alpha = 1, 2.
\label{EqzPolarRepresentation}
\end{equation}

Integration over the $\theta_i$ fields eliminates rapidly oscillating phase dependent terms,
such as the second term in (\ref{EqA}). We  show in the Appendix that the same happens to the
second term in the second line in (\ref{EqH21}). The effective lagrangian density, therefore, becomes
\begin{eqnarray}
\mathcal{ L }_{  \mbox{\scriptsize{eff}}}
& = &
 \psi^{ \dagger }  \left[ i\gamma^0\gamma^\mu\partial_\mu -\mu\right] \psi
+
2 \rho_{ s } |  D_\mu z_i  |^2
%\nonumber \\
%& &
+
 \psi^{ \dagger }_\alpha  \gamma^0\gamma^\mu  \sigma^z_{\alpha\beta} \psi_\beta A_\mu
\nonumber \\
& &
 +  \frac{a^4 J_{ \mbox{\scriptsize{K}} }^2 }{ 2 \chi_{ \perp} } {\bf  s} \cdot {\bf  s }
\, .
\label{EqH22}
\end{eqnarray}

Observe that the magnetic interaction between the itinerant electrons and the localized spins
manifests ultimately as a gauge coupling between the electrons and the Schwinger boson fields,
mediated by the CP$^{1} $  vector field $A_\mu$, which becomes a gauge field.
Indeed, (\ref{EqH22}) is invariant under the gauge transformation
\begin{eqnarray}
\psi & \rightarrow & e^{i \Lambda} \psi
\, ,
\nonumber \\
\theta_i & \rightarrow & \theta_i + \Lambda
\, ,
\nonumber \\
A_\mu & \rightarrow & A_\mu - \partial_\mu \Lambda
\, .
\label{3}
\end{eqnarray}

Our aim is to obtain the net effective interaction among the conduction electrons, associated to the fermion fields
$\psi$. For this purpose we are going to functionally integrate over the  CP$^{1} $ fields, in order to derive
the resulting effective interaction. Before doing that, however we shall
express the effective lagrangian in an explicitly gauge invariant way.

We first introduce the gauge invariant phase-fields \cite{ijmp}
\begin{equation}
\chi_i = \theta_i + \frac{\partial_\mu A^\mu}{\Box} ,
\label{4}
\end{equation}
which are clearly invariant under (\ref{3}). In (\ref{4}),
 $1/\Box$ is the Green function of the $\Box =  \partial_\mu\partial^\mu$ operator.

We can now re-write the effective lagrangian density (\ref{EqH22})
in an explicitly gauge invariant form given by \cite{ijmp}
\begin{eqnarray}
\mathcal{ L }_{  \mbox{\scriptsize{eff}}}
& = &
 \psi^{ \dagger }  \left[ i\gamma^0\gamma^\mu\partial_\mu -\mu\right] \psi
+
\frac{1}{2} \sum_{i=1}^2\ \rho_i^2\ \partial_\mu \chi_i \partial^\mu \chi_i
%\nonumber \\
%& &
%\hspace{-1.8cm}
+ \frac{1}{4} F_{\mu\nu}\left [\frac{4 \rho_s}{-\Box} \right ] F^{\mu\nu}
\nonumber \\
& &
+ \psi^{ \dagger }_\alpha  \gamma^0\gamma^\mu  \sigma^z_{\alpha\beta} \psi_\beta A_\mu
%\nonumber \\
%& &
+ \frac{a^4 J_{ \mbox{\scriptsize{K}} }^2 }{ 2 \chi_{ \perp} } {\bf  s}
\cdot {\bf  s }
\, .
\label{5}
\end{eqnarray}
where we have used the constant $\rho_i$ approximation for $i=1,2$
and  $F_{\mu\nu}= \partial_\mu A_\nu - \partial_\nu A_\mu $.

%%%%%%%%%%%%%%%%%%%%%%%%%%%%
\section{The Effective Electron Interaction}

Let us now perform the functional integration over the bosonic fields
$\rho$, $\chi$ and $A_\mu$. This will ultimately generate the final
effective interaction among the conduction electrons.

We shall adopt the constant $\rho_i$  ($i=1,2$)
approximation for performing the functional integration over the Schwinger boson fields, namely,
$\rho_i$'s and $\chi_i$'s. This approximation usually reproduces the physical situation found in many materials. It implies that integration over the $z_i$ fields would just yield a trivial multiplicative constant in the partition function.
The nontrivial interaction effect comes from integration over the gauge field $A_\mu$. This can be easily performed
given the quadratic dependence of (\ref{5}) in this field.

The resulting effective lagrangian density for the conduction electrons is
\begin{eqnarray}
\mathcal{ L }_{  \mbox{\scriptsize{eff}},\psi}
& = &
\psi^{ \dagger }  \left[ i\gamma^0\gamma^\mu\partial_\mu -\mu\right] \psi
%\nonumber \\
%& &
+
\frac{1}{8 \rho_s}
\left( \psi^{ \dagger }_\alpha  \gamma^0\gamma^\mu  \sigma^z_{\alpha\beta} \psi_\beta \right)
\left( \psi^{ \dagger }_\alpha  \gamma^0\gamma_\mu  \sigma^z_{\alpha\beta} \psi_\beta \right)
\nonumber \\
& &
+  \frac{a^4 J_{ \mbox{\scriptsize{K}} }^2 }{ 2 \chi_{ \perp} } {\bf  s} \cdot {\bf  s }\
\, .
\label{6}
\end{eqnarray}

Explicitly writing the components of the Dirac field we may, after some algebra, express the effective
interaction term above as

\begin{eqnarray}
\mathcal{ L }_{  \mbox{\scriptsize{I}},\psi}
& =  &
\frac{1}{4 \rho_s}
\left(
\psi^\dag_{1\uparrow } \ \psi^\dag_{2\downarrow } + \psi^\dag_{2\uparrow } \ \psi^\dag_{1\downarrow }
\right)
\left(
\psi_{2\downarrow } \ \psi_{1\uparrow }
+ \psi_{1\downarrow } \ \psi_{2\uparrow }
\right )
%\nonumber \\
%& &
\nonumber \\
& &
+
\frac{1}{8 \rho_s}\left [ \left ( \bar\psi_\sigma \psi_\sigma \right )^2
-
 \left ( \bar\psi_\sigma \gamma^0 \psi_\sigma \right )^2\right ]
\nonumber \\
& &
+  \frac{ a^2 J_{ \mbox{\scriptsize{K}} }^2 }{ 8 J }  {\bf  s} \cdot {\bf  s } - \frac{a^2 J }{2  }  s_z^2
\, ,
\label{7}
\end{eqnarray}
where $ {\bf  s}$, the spin of the itinerant electrons, is given by (\ref{Eqs}).

 The first term above is a superconducting, $ s $-wave BCS-type interaction,
the second term is a Nambu-Jona-Lasinio-type interaction,
that tends to produce an insulating charge-gapped phase showing an excitonic condensate.
The remaining terms  in (\ref{7}) correspond to an anisotropic spin-spin interaction given by

\begin{eqnarray}
\mathcal{ L }_{  \mbox{\scriptsize{s}},\psi} =
\frac{ a^2}{ 8 J }\left[ \left( J_{ \mbox{\scriptsize{K}} }^2 - 4 J^2 \right) s_z^2
+   J_{ \mbox{\scriptsize{K}} }^2  {\bf  s}_\perp \cdot {\bf  s }_\perp \right]
\, ,
\label{7aa}
\end{eqnarray}
where  $ {\bf  s}_\perp$ contains the $X,Y$ components of the itinerant electron spins.
Notice that the spin-spin interaction term above has the same symmetry as the XXZ Model.

%%%%%%%%%%%%%%%%%%%%%%%%%%%%
\section{The Phase Diagram}
\subsection{The Free Energy}

Let us introduce  the Hubbard-Stratonovitch (HS) auxiliary fields
\begin{eqnarray}
\Delta & = & \lambda_{ \mbox{\scriptsize{sc}} } \left(\psi_{2\downarrow } \ \psi_{1\uparrow } + \psi_{1\downarrow } \ \psi_{2\uparrow } \right)
\, ,
\nonumber \\
M & = &  \lambda_{ \mbox{\scriptsize{exc}} } \bar\psi_\sigma \psi_\sigma
\, ,
\nonumber \\
s_{ z } & =  &  \lambda_{ \mbox{\scriptsize{m}}}  \psi^{ \dagger }_\gamma \sigma_{\gamma \delta}^{ z } \psi_\delta
\, ,
\label{EqHubbard-Stratonovitch}
\end{eqnarray}
whose vacuum expectation values are, respectively, the relevant order parameters for superconducting, insulating and
magnetic orderings.
In terms of these fields, following the usual HS procedure, we replace the
fermionic quartic interactions in (\ref{7}) by quadratic and trilinear terms, namely

\begin{equation}
\mathcal{ L }_{  \mbox{\scriptsize{HS}}}
=
\int
\frac{ d^{ 2 } k }{ \left( 2 \pi \right)^{ 2 } }
\,
\Phi^{ \dagger } ( k )
\, \mathcal{A} \,
\Phi( k )
- \frac{ | \Delta |^{ 2 } }{ \lambda_{ \mbox{\scriptsize{sc}} } }
- \frac{ M^{ 2 } }{ \lambda_{ \mbox{\scriptsize{exc}} } }
- \frac{ s_{z }^{ 2 } }{ \lambda_{ \mbox{\scriptsize{m}} } }
\, ,
\label{EqHMF}
\end{equation}
where the Nambu field $ \Phi( k ) $ is given by
$ \Phi^{ \dagger }( k ) =
\left(
\psi^{ \dagger }_{ 1 \uparrow }(  k ) \;
\psi^{ \dagger }_{ 2 \uparrow }(  k ) \;
\psi_{ 1 \downarrow }( -k ) \;
\psi_{ 2\downarrow }( -k ) \;
\right) $
and
\begin{eqnarray}
\mathcal{A} =
%\begin{pmatrix}
\left(
 \begin{array}{cccc}
-\mu - M - s_{ z } & v_{ F } k_{ - } &  0  & -\Delta \\
 v_{ F } k_{ + } &  -\mu + M - s_{ z } & -\Delta  & 0 \\
0 & -\Delta^{ * }  & \mu + M - s_{ z } & -v_{ F } k_{ + }  \\
 -\Delta^{ * }  & 0 &  - v_{ F } k_{ - } & \mu - M - s_{ z }
\end{array}
\right)
%\end{pmatrix}
\, ,
\label{EqMatrixA}
\end{eqnarray}

In the previous expression $ v_{ F } $ denotes the Fermi velocity,
$ k_{ \pm } = k_{ y } \pm i k_{ x } $
and the effective couplings are given by

$$ \lambda_{ \mbox{\scriptsize{sc}} }
= \frac{1}{  4 \rho_{ s }  } \ \ \ \ \,\ \ \ \
 \lambda_{ \mbox{\scriptsize{exc}} }
=\frac{1}{  8 \rho_{ s }  } \ \ \ \ \,\ \ \ \
 \lambda_{ \mbox{\scriptsize{m}} }
= \frac{ a^2}{ 8 J } \left( J_{ \mbox{\scriptsize{K}} }^2 - 4 J^2 \right)
$$

The mean-field approach consists in replacing the HS-fields by their vacuum
expectation values, whereupon the fermionic integrals become quadratic and the
different order parameters are introduced.

Two comments are in order at this point.
Firstly, the term $\left ( \bar\psi_\sigma \gamma^0 \psi_\sigma \right )^2$ after the HS-transformation
and mean-field approximation, just produces a shift in the chemical potential. Secondly,
we have chosen the direction of the spin (vector) order parameter along the z-axis. It follows that
within the mean-field approach the vacuum expectation value of the transverse components of ${\bf s}$
vanish:  $\langle {\bf  s}_\perp \rangle = 0$.
As a consequence of the above remarks we have neglected the terms containing $\bar\psi_\sigma \gamma^0 \psi_\sigma $~\cite{Nambu1961}
and ${\bf  s}_\perp$ from (\ref{7}) and (\ref{7aa}).

Now, we proceed to the evaluation of the partition function as a path integral in the complex time representation
and integrate it over the fermionic fields, thereby obtaining
the free energy (effective potential)~\cite{Nunes2010} as a function of the different order parameters, namely

\begin{eqnarray}
V_{ \mbox{\scriptsize{eff}} }
& = &
-
T
\sum_{ n = -\infty }^{ \infty }
\int \frac{ d^2 k }{ (2 \pi )^2 }
\,
\log
\left\{
\frac
{
\prod_{ j = 1}^{ 4 }
\left[ \,
i \omega_n -  E_j
\, \right]
}
{
\prod_{ j = 1}^{ 4 }
\left[ \,
i \omega_n -  E_j( M = 0, \Delta = 0 )
\, \right]
}
\right\}
\nonumber \\
& &
+ \frac{ | \Delta |^2 }{ \lambda_{ \mbox{\scriptsize{sc}} } }
+ \frac{ M^2 }{ \lambda_{ \mbox{\scriptsize{exc}} } }
+ \frac{ s_{z}^2 }{ \lambda_{ \mbox{\scriptsize{mo}} } }
\, ,
\label{EqVeff}
\end{eqnarray}
where $ \omega_n =  ( 2 n + 1 ) \pi T $ are the Matsubara frequencies for fermions
(with the Boltzmann constant $k_B = 1 $) and the four values of $ E_ j $'s are
\begin{equation}
E_j =
\pm
\epsilon_{ \pm } ( v_{ F } k )
- s_{ z }
\label{Ej}
\, ,
\end{equation}
with $ j = 1, \cdots,  4 $ and

\begin{equation}
\epsilon_{ \pm } ( x)
=
\sqrt{
\left[
\sqrt{ x^{ 2 } + M^{ 2 } + \left( \frac{ | \Delta | M }{ \mu } \right)^{ 2 } } \pm \mu
\right]^{ 2 }
+
| \Delta |^{ 2 } - \left( \frac{ | \Delta | M }{ \mu } \right)^{ 2 }
}
\, .
\label{EqEpsilon}
\end{equation}

\subsection{ Competing Phases }

Let us analyze the possible phases that derive from the free energy.
The appearance of superconductivity, excitonic condensate or magnetic order is associated, respectively,
to the existence of nonzero solutions for $ \Delta $, $ M $ or $s_z$, which minimize the free energy.
In what follows we consider different regimes of couplings.

\subsubsection{ $  \lambda_{ \mbox{\scriptsize{mo}} } \gg \lambda_{ \mbox{\scriptsize{sc}} }, \lambda_{ \mbox{\scriptsize{exc}} } $ }

In this regime we only take into account the magnetic interaction and
the system is reduced to a set on non-interacting relativistic electrons with a Zeeman splitting.
Through the mean-field approach proposed in (\ref{EqHubbard-Stratonovitch}),
the z-component spin part in (\ref{7}) is reduced to a term
that is proportional to $ s_{z  }  = n_{ \uparrow } - n_{ \downarrow } $.
This is added to the term containing the chemical potential in (\ref{6}),
splitting the chemical potential into two contributions.
This indicates that the energies required do add or subtract fermions with spins $ \uparrow $ or $ \downarrow $ are different,
allowing the possibility of spin population imbalance, which leads to a magnetically ordered phase.
This case has been previously investigated~\cite{Caldas2009}.

\subsubsection{ $  \lambda_{ \mbox{\scriptsize{sc}} } \gg \lambda_{ \mbox{\scriptsize{mo}} }, \lambda_{ \mbox{\scriptsize{exc}} } $ }

Now, only the superconducting interaction is important.
In such case there is threshold for the superconducting interaction coupling
$\lambda_{ \mbox{\scriptsize{sc}} }\lambda_{ \mbox{\scriptsize{sc}} }$, below which
the system is in the normal phase for $ \mu = 0 $, even at $T=0$
This quantum critical point $ $
occurs at~\cite{Marino2006}.
$ \lambda_{ \mbox{\scriptsize{sc}} }^c  = 2 \pi v^{ 2 }_{ F } / \Lambda $,
where $ \Lambda $ is the  ultraviolet cutoff, which is naturally provided by the subjacent lattice structure.

As charge carriers are added to the system by doping,
the chemical potential becomes non-zero and, as $  \mu  $ increases, the system becomes superconducting. Eventually
 $  \mu  $ reaches an optimum doping for which $ \Delta $ is maximum~\cite{Nunes2010}
with the order parameter displaying a dome-shaped plot,
which vanishes exponentially as $ \mu \rightarrow 0 $.

\subsubsection{ $  \lambda_{ \mbox{\scriptsize{sc}} },  \lambda_{ \mbox{\scriptsize{exc}} } \gg \lambda_{ \mbox{\scriptsize{mo}} } $ }

When both the superconducting and the excitonic interactions are taken into account,
but the magnetic term is disregarded
one can show that the effective potential at $ T = 0 $
in (\ref{EqVeff}) becomes ( except for an additive term that does not depend on the order parameters )
\begin{equation}
V_{ \mbox{\scriptsize{eff}} }
=
\frac{ \Lambda^{ 3 } }{ \alpha }
\left(
\frac{ | \Delta |^{ 2 } }{ \tilde{\lambda}_{ \mbox{\scriptsize{sc}} } }
+
\frac{ M^{ 2 } }{ \tilde{\lambda}_{ \mbox{\scriptsize{exc}} } }
-
\sum_{ j = \pm 1 } \int_{ 0 }^{ 1 } d y \, \epsilon_{ j } ( y )
\right)
\, ,
\label{EqVeff_T0_Sz0}
\end{equation}
where $ \epsilon_{ \pm } ( y ) $ is given by (\ref{EqEpsilon}),
with all the energy quantities given in units of $ \Lambda $
and the interaction strengths given by the dimensionless couplings
$ \tilde{ \lambda }_{ \mbox{\scriptsize{sc}} }
\equiv
\lambda_{ \mbox{\scriptsize{sc}} } /\lambda_{ \mbox{\scriptsize{sc}} }^c  =  2 \tilde{ \lambda }_{ \mbox{\scriptsize{exc}} } $.
Notice that the ratio  between the SC and excitonic couplings is fixed and equal to 2.

Numerical results for the minima of (\ref{EqVeff_T0_Sz0})
present no excitonic condensate for any value of chemical potential.
On the other hand, the superconducting phase diagram
is similar to the one for  $ \lambda_{ \mbox{\scriptsize{exc}} } = 0 $,
where there is an optimum doping as charge carriers are added to the system
and $ \Delta $ displays a dome-shaped plot.

%\subsubsection{ $ T \neq 0 $ }

For finite temperature, since we are looking for the minima of (\ref{EqVeff}),
we take the derivative of $ V_{\rm eff} $ with respect to $ \Delta $ and $ M $
after the summation over the Matsubara frequencies.
The nonzero solutions of the order parameters provide the following two coupled equations,
\begin{eqnarray}
\frac{ 1 }{ \lambda_{ \mbox{\scriptsize{sc}} } }
& = &
\,
\sum_{ j = \pm 1 }
\int \frac{ d^{ 2}k }{ \left( 2 \pi \right)^{ 2 } }
\,
\frac{ 1  }{ 2 \Delta }
\frac{ \partial \epsilon_{ j }  }{ \partial \Delta }
\tanh\left( \frac{ \beta }{ 2 } \epsilon_{ j } \right)
\, ,
\label{EqDeltaEquation}
\\
\frac{ 1 }{ \lambda_{ \mbox{\scriptsize{exc}} } }
& = &
\,
\sum_{ j = \pm 1 }
\int \frac{ d^{ 2}k }{ \left( 2 \pi \right)^{ 2 } }
\,
\frac{ 1  }{ 2 M }
\frac{ \partial \epsilon_{ j }  }{ \partial M }
\tanh\left( \frac{ \beta }{ 2 } \epsilon_{ j } \right)
\, ,
\label{EqMEquation}
\end{eqnarray}
where $ \beta = T^{ -1 } $.

The critical temperatures are defined as the temperatures
below which the order parameter develops a nonzero value.
Hence, we simply make
$ \Delta =  0 $ at $ T = T_{ \Delta } $ and $ M = 0 $ at $ T = T_{ M } $
in (\ref{EqDeltaEquation}) and (\ref{EqMEquation}) respectively
and we find the self-consistent equations for the critical temperatures.

Remember that the Coleman-Mermin-Wagner-Hohenberg theorem~\cite{MermimWagner1966}
rules out any finite temperature phase transition for a 2D system, therefore, as usual
$ T_{ \Delta } $ might be regarded as the temperature for the onset of the
Berezinskii-Kosterlitz-Thouless (BKT) phase transition~\cite{Berezinskii1973},
below which phase coherence is combined with a nonzero $ |\Delta |$. This mechanism is possible
because the SC order parameter is complex. It does no apply, in particular, to the real excitonic order
parameter $ M $, which results zero at finite temperature, in agreement with the theorem.
Nevertheless, for a large number of stacks of such planar system, $T_M$ becomes nonzero \cite{SSC}
and the calculated temperatures are the actual phase transition temperatures,
which might be realized experimentally.

%%%%%%%%%%%%%%%%%%%%%%%%%%%%
\begin{figure}[ht]
\centerline
{\includegraphics[clip,angle=90,width=0.75\textwidth]{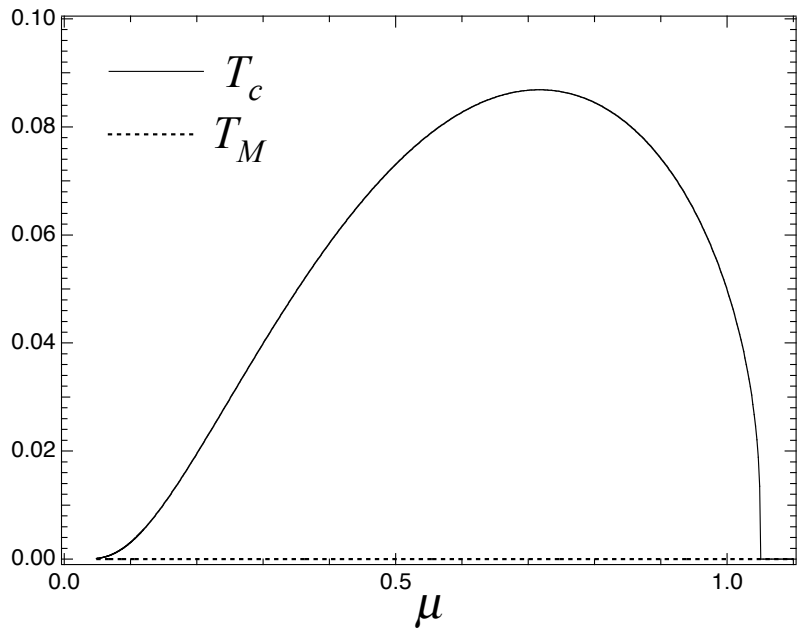}}
\caption{$ T_{ \Delta } $ and $ T_{ M } $ as a function of $ \mu $ for $  \tilde{ \lambda }  = 0.8 $.}
\label{FigTDTm_X_mu}
\end{figure}
%%%%%%%%%%%%%%%%%%%%%%%%%%%%%%%
Our results for $  \tilde{ \lambda }_{ \mbox{\scriptsize{sc}} } = 0.8 < 1 $
can bee seen in Fig.~\ref{FigTDTm_X_mu}.
For this particular choice of parameter,
there is no superconductivity at $ \mu = 0 $,
since the interaction strength is below the quantum critical point,
$ \lambda_{ \mbox{\scriptsize{sc}} } < \lambda_{ \mbox{\scriptsize{sc}} }^{ c } $.
Moreover, there is no  $ T_{ M } $ for any range of chemical potential,
which is in agreement with the CMWH theorem.

As $ \mu $ increases,
the model displays a finite superconducting critical temperature, $ T_{ \Delta } $.
This is a manifestation of the Cooper's theorem,
which states that superconductivity always occurs below a finite temperature when a Fermi surface builds up.
Indeed, even when the interaction is not strong enough to produce superconductivity at half filling,
the system develops a superconducting state as charge carriers are added to the system and the chemical potential
becomes different from zero..
Moreover, as $ \mu $ increases, $ T_{ \Delta } $ displays a dome-shaped curve
reaching an optimal value for the critical temperature and decreasing for higher values of chemical potential.
This result qualitatively reproduces the same features experimentally observed in several compounds,
like 122 pnictides, cuprate superconductors and other two-dimensional systems with Dirac fermions~\cite{Smith2009,Nunes2005}.

The complete analysis of the phase diagram
when the superconducting, excitonic and magnetic interaction terms are taken into account
can be rather lengthy, since it introduces an extra free parameter. It is not necessary for our
purposes here. Indeed, the main scope of this paper is
to provide an example where the common origin of different competing interactions could be traced
back. For this, the analysis of the phase diagram presented here is sufficient.

%%%%%%%%%%%%%%%%%%%%%%%%%%%%
\section{Conclusions}\label{Conclusions}

We have provided a concrete example of a model containing both localized spins and itinerant electrons, where the
different effective competing interactions occurring among these can be traced back to the original purely magnetic interactions. In other words,
we have shown that the AF Heisenberg-Kondo lattice model, which only contains magnetic interactions, both among the localized spins and between those and the itinerant electrons (Kondo interaction) ultimately will generate among the latter a superconducting BCS-like
interaction, an excitonic Nambu-Jona-Lasinio-like interaction and a magnetic, spin-spin interaction. These interactions will compete,
depending on the temperature, chemical potential and coupling constants, yielding a rich phase diagram.
This is analyzed in three different regimes of coupling parameters at $T=0$ and in the regime where the magnetic coupling
may be neglected at $T \neq 0$.

 A quantum critical point $ \lambda_{ \mbox{\scriptsize{sc}} }^{ c } $ exists for $T = \mu =0$, below which
the system is in the normal phase. As we dope the system and the chemical potential becomes
non-zero, a Fermi surface builds up and the quantum phase transition accordingly is  washed out
in agreement with Cooper's theorem.

A finite SC transition temperature was found, by means of the BKT mechanism, which
applies in the case of a complex order parameter. In the case of the excitonic condensate transition a
vanishing transition temperature was obtained in agreement with the CMWH theorem and the fact that the
corresponding order parameter is real.

In our model, superconductivity arises from a novel mechanism, which originates from
the purely magnetic interactions involving the localized spins and itinerant electrons of the
original system. Interestingly, the SC effective coupling parameter is inversely proportional to the
original AF Heisenberg coupling, suggesting that the onset of superconductivity is related with
a decrease of the  magnetic ordering. {\bf This is in agreement with the phenomenological
scenario for materials such as cuprates, pnictides and heavy-hermions}.
{\bf
This result is also consistent with DMRG calculations for the 1d Heisenberg-Kondo model \cite{Xavier2008}, which shows the development of a superconducting phase mediated by antiferromagnetic fluctuations. On the same token, recent mean field calculations for the 2d Kondo lattice \cite{Liu2012}, shows that, as an AFM Heisenberg exchange coupling is taken into account, singlet Cooper pair occurs among conduction electrons, leading to heavy fermion superconductivity. Hence, our derivation may provide an analytical explanation for those results without the resort of any approximation, which is a very interesting result.
}

The dome-shaped dependence of the temperature on the chemical potential suggests,
on the other hand, that Dirac electrons might play a relevant role
in some of those strongly correlated compounds
and similar conclusions have been withdrawn from simulations for pairing
in ultracold atoms in hexagonal arrangement~\cite{Lim2009}.

%%%%%%%%%%%%%%%%%%%%%%%%%%%%
\ack
This work has been supported by CNPq, FAPERJ and FAPEMIG.
We would like to thank H. Caldas, A. L. Mota, R. L. S. Farias and M. B. Silva Neto for discussions.

%%%%%%%%%%%%%%%%%%%%%%%%%%%%
%Begging of Appendices
%%%%%%%%%%%%%%%%%%%%%%%%%%%%
%\appendix
\section{Appendix}\label{Appendix}

Let us show here that the second term in the second line in (\ref{EqH21}) only contains rapidly oscillating phase dependent terms,
which are, therefore, eliminated through functional integration over the phase fields.

We may write this term as
\begin{eqnarray}
\left (  \psi^\dagger \vec{ \sigma } \psi  \right) \cdot \left( {\bf n } \times \partial_{ \tau } {\bf n } \right )
& = &
\psi^\dagger_\alpha  \psi_\beta  z^*_\mu z_\nu \partial_\tau \left (z^*_\lambda z_\rho \right )
%\nonumber \\
%& &
\times  \epsilon^{ijk} \sigma^i_{\alpha\beta} \sigma^j_{\mu\nu}  \sigma^k_{\lambda\rho}
\, ,
\label{7a}
\end{eqnarray}
where we have used (\ref{Eqz1z2}).

Now, we have the following identity for Pauli matrices
\begin{eqnarray}
\sigma^i_{\alpha\beta} \sigma^j_{\mu\nu}
& = &
\frac{\delta^{ij}}{3}
\left [  2 \delta_{\alpha\nu} \delta_{\beta\mu}- \delta_{\alpha\beta} \delta_{\mu\nu} \right]
%\nonumber \\
%& &
+ i \epsilon^{ijk} \left [ \delta_{\beta\mu} \sigma^k_{\alpha\nu}
- \delta_{\alpha\mu} \sigma^k_{\beta\nu}\right ]
\, .
\label{8}
\end{eqnarray}

Inserting (\ref{8}) in (\ref{7a}), we immediately obtain
\begin{equation}
 {\bf s} \cdot \left( {\bf n } \times \partial_{ \tau } {\bf n } \right )
  =
  \psi^\dagger_\alpha  \psi_\alpha  z^*_\mu z_\mu \partial_\tau \left (z^*_\lambda z_\lambda \right )
+ pdt ,
\label{9}
\end{equation}
where $pdt$ stands for ``phase dependent terms''.

The first term on the r.h.s. above vanishes because $ z^*_\lambda z_\lambda = 1$ and, therefore, only phase
dependent terms are left, as we have asserted.

%%%%%%%%%%%%%%%%%%%%%%%%%%%%
\bibliography{apssamp}

\end{document}